\newcommand{\be}{\begin{equation}}
\newcommand{\ee}{\end{equation}}
\newcommand{\bea}{\begin{eqnarray}}
\newcommand{\eea}{\end{eqnarray}}
\newcommand{\bean}{\begin{eqnarray*}}

\newcommand{\eean}{\end{eqnarray*}}
\font\upright=cmu10 scaled\magstep1
\font\sans=cmss10
\newcommand{\ssf}{\sans}
\newcommand{\stroke}{\vrule height8pt width0.4pt depth-0.1pt}
\newcommand{\Z}{\hbox{\upright\rlap{\ssf Z}\kern 2.7pt {\ssf Z}}}

\newcommand{\C}{{\rlap{\rlap{C}\kern 3.8pt\stroke}\phantom{C}}}
\newcommand{\R}{\hbox{\upright\rlap{I}\kern 1.7pt R}}
\newcommand{\CP}{\C{\upright\rlap{I}\kern 1.5pt  P}}
\newcommand{\PP}{\hbox{\upright\rlap{I}\kern 1.5pt  P}}

\newcommand{\identity}{{\upright\rlap{1}\kern 2.0pt 1}}

\newcommand{\HH}{\mbox{\hbox{\upright\rlap{I}\kern 1.7pt H}}}

\newcommand{\fr}{\frac}

\newcommand{\ra}{\rightarrow}
\newcommand{\al}{\alpha}
\newcommand{\pr}{\partial}
\newcommand{\hs}{\hspace{5mm}}

\newcommand{\acc}{\\[3mm]}

\newcommand{\ie}{{\it ie }}

\newcommand{\prm}{{\prime}}

\input{epsf}

\documentstyle[12pt,a4wide]{article}

\begin{document}
\title{\vskip -70pt
\begin{flushright}
{\normalsize UKC/IMS/00/12} \\
\end{flushright}\vskip 50pt
{\bf \large \bf
Kink Dynamics in a Lattice Model with Long-Range Interactions}}
\author{T. Ioannidou$^1$\thanks{Permanent Address: Institute of
Mathematics, University of Kent, Canterbury CT2 7NF, UK}\,,
J. Pouget$^1$ and  E. Aifantis$^2$\thanks{Center for Mechanics of
Materials and Instabilities, Michigan Technological University, Houghton, MI
 49931, USA}
\\[10pt]
$^1${\normalsize  {\sl Laboratoire de Mod\'elisation en 
M\'ecanique (associ\'e au CNRS),}}\\
{\normalsize  {\sl Universit\'e Pierre et Marie Curie, 
Tour 66, 4 place Jussieu,}}\\
{\normalsize  {\sl 75252 Paris C\'edex 05, France}}\\[10pt]
$^2${\normalsize  {\sl Laboratory of Mechanics and Materials, 
Polytechnic School,}}\\
{\normalsize  {\sl Aristotle University of Thessaloniki, 
54006, Thessaloniki, Greece}}} \date{}

\maketitle

\begin{abstract}
This paper proposes a one-dimensional lattice model with long-range
interactions which, in the continuum, keeps its nonlocal behaviour.
In fact, the long-time evolution of the localized waves is governed
 by an asymptotic equation of the Benjamin-Ono type  and allows 
the explicit construction  of moving kinks on the lattice.
The long-range particle interaction coefficients on the lattice 
are determined by the Benjamin-Ono equation.

\end{abstract}

\section{Introduction}

This paper is concerned with the effects of long range interactions
in the behaviour of solitons (kinks)  in a lattice system.
In particular, we show that the introduced lattice model 
asymptotically leads to a  nonlocal continuum equation. 
By nonlocal we mean the model where the localized solution 
is of order of the scale parameter ({\it ie} parameter which has
 the dimension of length), and thus it is physically acceptable 
to consider wavelengths comparable with the scale parameter.
Its equation of motion  contains
integral, integro-differential or finite difference operators in the spatial
variables, while the wave propagation velocity depends on the wavelength.
In the present model there are two characteristic lengths: (a) the lattice
spacing and (b) the radius of particle interaction range.

The nonlocal models can be divided into two classes: discrete and continuous. 
For specific cases, it is simpler and easier to use the quasicontinuum 
description of the discrete medium.
Its essence is an interpolation of functions of discrete argument by a
special class of analytic functions in such a way that a
correspondence condition between the quasicontinuum and the 
discrete medium is fulfilled.
The advantages of such an approach is the description of discrete
and continuum media within a unified formalism and their 
 correct generalization.
In that case the quasicontinuum model is also applicable to macrosystems.
Analytic solutions valid for wide nonlinear excitations
 have been obtained using the continuum approximation in \cite{PPF,FPR} and
the quasicontinuum approximation in \cite{Co,R1,R2}.
In particular in \cite{PPF,HMB}, it has been shown that very narrow 
soliton-like excitations can propagate without any energy loss 
due to discreteness effects.

The lattice model we study, possesses an inherent
nonlocal behaviour, \ie the motion of a particle at
node $n$ depends on the motion of its neighbours.
Such models for different kinds of  particle interactions  have
been studied in a series of papers \cite{A,RF,MMWI}.
So far the classical procedure for studying them, at the continuum
limit, is to assume  slowly varying fields and expand 
them using Taylor's series.
Then, approximate  partial differential equations for the
continuous fields were obtained.

In this paper, we show how to keep the nonlocal behaviour 
of the lattice model  at the continuum level,  or in other words, 
how to describe the nonlocality at the continuum level.
This transition from the continuum to the discrete
 system with long-range interactions and vice-versa,
 is straightforward due to the quasicontinuum approximation.
The idea of preserving the nonlocality at the continuum limit 
 occurs in condensed-matter physics to describe phase transition in
crystals, nonlinear waves in crystals and biological molecules.
For example, modified versions of the discrete nonlinear Schr\"odinger
 equation  which keep the nonlocality at the continuum and
describe problems in biomolecules have been studied in the
past \cite{Cet1,Cet2}.

It is worthwhile mentioning some interesting studies of anharmonic atomic 
chain including long-range interactions due to large Coulomb coupling 
between particles  \cite{KK,GFNM,BRP,CGV}.
 The long-range potential is of the inverse  power-type or Kac-Baker form.
 In the case of inverse power type of the fourth-order, the continuum limit 
yields an integro-differential equation involving 
a Hilbert transform \cite{CGV}.
Using a perturbative technique the authors derive a 
mixed modified Korteweg-de-Vries with Benjamin-Ono equations \cite{CGV}, which is 
close to the result presented in the following.

The  paper is organised as follows:
In Section 2 we introduce the lattice model; while in Section 3, 
we deal with its continuum approximation and show that it
leads to the  Benjamin-Ono equation for long-time evolution.
In Section 4, we derive the type of long-range particle interactions on the
lattice using the fact that the lattice model asymptotically leads to
 the Benjamin-Ono equation.
Finally, numerical simulations of the discrete model based on the 
Benjamin-Ono soliton are given and discussed in Section 5.

\section{The Lattice Model}

We consider a one-dimensional lattice with mass $M$, lattice spacing $b$ and
long-range interactions.
The Lagrangian  of the system is
\be
{\cal L}=\sum_{n}\fr{1}{2}M\dot{u}_n^2-
\sum_{nn^{\prime}}\fr{1}{4}\Psi(n-n^{\prime})\,\left(u_n
-u_{n^{\prime}}\right)^2
-\sum_{n}\fr{1}{3}k_3(u_{n+1}-u_n)^3.
\label{ldis}
\ee
Figure 1 presents a one-dimensional chain of
masses and springs with second neighbour interactions.

\begin{figure}
\unitlength1cm
\begin{picture}(5,5)
\put(2.85,2.25){$b$}
\put(4.75,3.45){$M$}
\put(13.55,5.45){$n+2$}
\put(10.65,5.45){$n+1$}
\put(7.75,5.45){$n$}
\put(4.45,5.45){$n-1$}
\put(1.15,5.45){$n-2$}
\put(6.7,2.25){$\Psi(1)$}
\put(5.25,2.25){$k_3$}
\put(8.3,0.65){$\Psi(2)$}
\epsfxsize=10cm
\epsffile{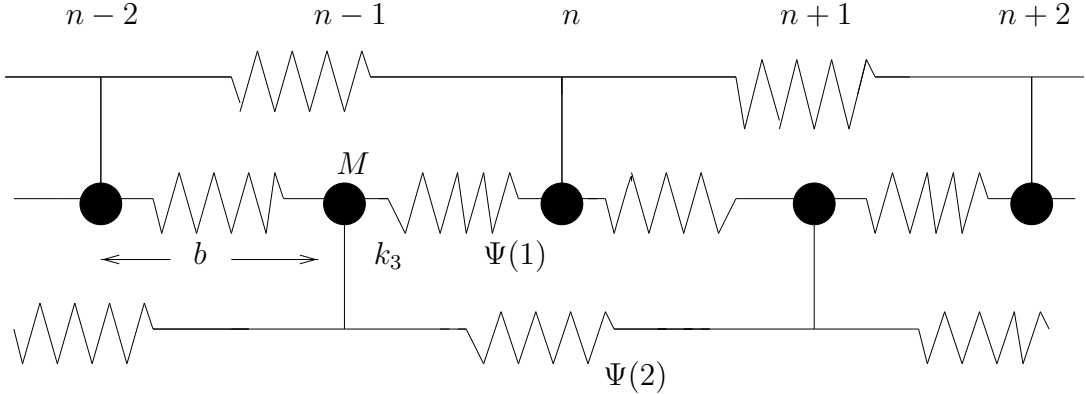}
\end{picture}
\caption{One-dimensional atomic chain of masses and springs with first and
second neighbour interactions.}
\end{figure}

The equations of motion obtained from the variation of (\ref{ldis}) are
\be
M\ddot{u}_n=\sum_{m=1}^{{\cal M}}\Psi(m)\,(u_{n+m}+u_{n-m}-2u_n)
+k_3\left[(u_{n+1}-u_n)^2-(u_n-u_{n-1})^2\right],
\label{dis}
\ee
where $k_3$ is a positive constant ($k_3<1$).
$\Psi(n)$ are the interaction coefficients of the particles,
related  to the force constants $\Phi(n)$ (see equation(\ref{Psi}))
 which determine the effective
characteristics of the elastic bonds between the particles.
The lattice parameters $\Psi(n)$ can be obtained from the 
interaction potential between particles or atomic bonds in
crystals (same as for $\Phi(n)$); e.g. the Born-Meyer potential, the Morse
 potential, the Lennard-Jones potential, the covalent potential, etc.

Here, we consider the chain to be homogeneous and so,
 the unique definition of  $\Phi(n)$  leads to the 
identity $\Phi(n)=\Phi(-n)$.
If $\Phi(n)$  differs from zero for $|n| \leq {\cal M}$,
 then every particle interacts with ${\cal M}$ neighbours to the
right and to the left.
The case ${\cal M}=1$ corresponds to the simplest model of interaction 
between first nearest neighbours; while
in real mechanical systems the action at a distance is always restricted,
\ie ${\cal M}$ is finite.

The distances between the particles do not change and therefore the forces
acting on them are equal to zero \cite{K}.
This leads to the following condition for the force constants 
\be
\Phi(0)=-\sum\nolimits'_{n}\Phi(n),
\label{Pcon}
\ee
where $\sum_{n}^\prm$ means summation over all $n\neq 0$.
This equation is deduced from the lattice energy invariance; \ie
the invariance by translation of the chain as a whole.

Assume now, that the particles interact nonlocally with measure
$\Psi(n)$. Then it can be shown that \cite{K}
\be
\Phi(n)=-\Psi(n), \hs \hs n\neq 0; \hs \hs \hs
\Phi(0)=\sum\nolimits'_{n}\Psi(n).
\label{Psi}
\ee
Note that for  real mechanical systems all $\Psi(n)>0$ and hence
$\Phi(0)>0$, $\Phi(n)<0$ ($n\neq 0$).
However, a system may be constructed, for which some particles would have
negative interaction coefficients; which is the case of our model 
described below.

\section{The Continuum Approximation}

The set of nonlinear differential equations for a discrete system is
generally complex and, quite often, its continuum approximation 
is considered (which is the case here).
By assuming that the displacements are slowly varying over the lattice
spacing, the continuum approximation may be derived as follows:

\subsection{The Nonlocal Model}

The Lagrangian (\ref{ldis}) in the form which is invariant
 with respect to $n$ and $x$ representations, can be written as 
\be
{\cal L}=\fr{1}{2}<\rho \dot{u}, \dot{u}>+\fr{1}{2}<u,\,\Phi u>
-\fr{1}{3}<\Delta u,\, k_3 \Delta u^2>,
\label{lcon}
\ee
where  $\rho$ is the mass density, while
 $u=u_n(t)$, $\Delta u=b^{-1}(u_{n+1}-u_n)$ and
$u=u(x,t)$, $\Delta u=u_{x}$ in the discrete and continuum space.

The various terms in (\ref{lcon})  for both the discrete and the 
continuum representations are given as follows
\begin{eqnarray}
\fr{1}{2}<u,\,\Phi u>&=&\fr{1}{4}\sum_{n n^\prime}\Phi(n-n^\prime)
(u_n-u_{n^\prime})^2\nonumber \\
&=&\fr{1}{2} \int_{R^2}\Phi(x-x^\prime) \, u(x)
\, u(x^\prime)\, dx \,dx^\prime,\acc
\fr{1}{3}<\Delta u,\, k_3 \Delta u^2>&=&
\fr{1}{3}\sum_{n}k_3(u_{n+1}-u_n)^3\nonumber\\
&=&\fr{1}{3}\int_{R}k_3\,u_x^3\,dx.
\end{eqnarray}
The above forms contain elastic (long-range) interactions for both
 the discrete model and its quasicontinuum counterparts.

The equation of motion of the medium in the $(x,t)$ representation is
\be
\rho\ddot{u}(x,t)=-\int_R \Phi(x-x^\prm) u(x^\prm,t) \,dx^\prime
+k_3 (u_x^2)_{x}.
\label{cP}
\ee

The corresponding condition (\ref{Pcon})  in the $x$ representation 
becomes $\int_R \Phi(x-x^\prime)\, dx^\prm=0$ which implies that 
$\Phi(x-x^\prime)$ can be represented as
\be
\Phi(x-x^\prime) =\psi(x)\,\delta(x-x^\prime)-\Psi(x-x^\prime),
\ee
with $\Psi(x-x^\prime)=\Psi(x^\prm-x)$ and $\psi(x)\doteq\int
\Psi(x-x^\prime)\, dx^\prime$.

Its quasicontinuum counterparts are of the form
\be
\Phi(n-n^\prime) =\psi(n)\,\delta(n-n^\prime)-\Psi(n-n^\prime),
\label{Pdis}
\ee
with $\Psi(n-n^\prime)=\Psi(n^\prm-n)$ and $\psi(n)\doteq\sum_{n^\prm}
\Psi(n-n^\prime)$.
Thus $\Psi(n-n^\prime)$ and $\Psi(x-x^\prime)$ are the interaction
coefficients and the  interaction coefficient density of the model  
connecting the points ($n$, $n^\prm$)  and ($x$, $x^\prm$), respectively.

In the $x$ representation the force constant can be expressed  as \cite{K}
\be
\Phi(x-x^\prm)=\fr{\pr^2}{\pr x \pr x^\prm}c(x-x^\prm),
\label{22}
\ee
where $c(x-x^\prm)$ is the so-called kernel operator of elastic moduli.

Then, equation (\ref{cP}) after integration by parts
 becomes
\be
\rho\ddot{u}(x,t)=\fr{\pr}{\pr x}\int_R c(x-x^\prime)
\fr{\pr u(x^\prime)}{\pr x^\prime} \,dx^\prime
+k_3 (u_x^2)_{x},
\label{ro}
\ee
which is quite similar to the equation obtained from the
nonlocal elasticity of the one-dimensional linear case (\ie with the $k_3$
term vanishing) \cite{Kro1,Er}.

In this case, by using the kinematical definition for the strain
$\epsilon(x',t)=\pr u(x',t)/\pr x'$ and the nonlocal constitutive equation for
the stress $\sigma(x,t)=\int_{-l}^{l} c(x-x') \,\epsilon(x',t)\,dx'$
where $l$ is the interaction range length,  equation (\ref{ro}) can be
 recast into
\be
\rho\ddot{u}(x,t)=\sigma_x,
\label{24}
\ee
which is the equation of motion of a one-dimensional elastic medium.
[Note that, the integral over a finite radius of interaction defines 
the finiteness of action at a distance determined by this radius 
(that is, $l={\cal M}b)$. 
Specifically, $\Phi(y)$ is not zero for $y\in [-l ,l]$ and it equals to 
zero otherwise; while, if the chain is bounded within an interval $2L$ 
then $l<<L$.]

In this connection, equation (\ref{24}) is the starting 
point in continuum mechanics and expresses the balance of linear momentum.
It is endowed with appropriate constitutive equations for the stress leading 
to partial differential equations for the determination of the displacement
field.
For example, a simple linear elastic model of the form $\sigma=E \epsilon$ 
with $E$ denoting the Young modulus (identified as $c_0^2$ below) 
leads to the classical wave equation $u_{tt}=c^2 u_{xx}$ ($c^2\equiv E/\rho$).
If gradient-dependent nonlinear elasticity model is adopted \cite{TA} of the
form
$\sigma=f(\epsilon)-\tilde{c} \epsilon_{xx}$
the appropriate equation of motion reads $f(u_x)u_{xx}-\tilde{c}u_{xxxx}=\rho
u_{tt}$ which contain the $k_3 (u_x^2)_x$ term listed above, even though a
more interesting case results when $f^\prime(u_x)$ is non-monotonic.
Other types of such constitutive models and the corresponding equations of
motion are discussed in \cite{Aif}.
A case of particular interest in terms of its discrete model counterpart
arises when viscosity is considered.

For a simple Kelvin-Voigt model of the form $\sigma= E
\epsilon+\mu\dot{\epsilon}$ where $\mu$ is the viscosity and $\dot{\epsilon}$
the strain rate, the resultant equation of motion with $\lambda\equiv
\mu/\rho$ is $u_{tt}=c^2u_{xx}+\lambda \dot{u}_{xx}$, \ie the damped
 wave equation.
It is an open question to examine a Lagrangian-like lattice version of 
this model.

On the other hand, by letting
\be
c=c_0^2\,\delta(x-x^\prime)+c_1(x-x^\prime),
\label{c}
\ee
where $c$ corresponds to the local response with $c_0^2$ denoting the local
elastic modulus and $c_1$  describing the nonlocal behaviour  of the model,
 equation (\ref{ro}) simplifies to
\be
\rho\ddot{u}(x,t)=c_0^2\,u_{xx}+\fr{\pr}{\pr x}\int_R c_1(x-x^\prime)
\fr{\pr u(x^\prime)}{\pr x^\prime} \,dx^\prime
+k_3 (u_x^2)_x,
\label{nonl}
\ee
which is further examined below.

\subsection{The Long-Time Evolution of the Localized Wave}

In this section we investigate the asymptotic behaviour of equation 
(\ref{nonl}) for large times.
By assuming that the contribution of the nonlinear term is significant
throughout we rescale the nonlinear and the dispersive term 
by introducing a small parameter  $\varepsilon<1$, as follows
\be
2k_3=\gamma\,\varepsilon,\hs\hs\hs c_1=\varepsilon\,g.
\label{cod}
\ee
Here, $\gamma$ is the rescaled nonlinearity coefficient and $g$ is the
rescaled elastic kernel.
Then, for large times, the asymptotic expansion for the displacement field
can be expressed as
\be
u(x,t)=u_0(\xi,\tau)+\varepsilon\,u_1(x,t)+O(\varepsilon^2),
\label{exp}
\ee
where $\xi=x-vt$ is a shifted coordinate and $\tau=\varepsilon t$ is the 
slow time variable.

Using  equations (\ref{cod}) and (\ref{exp}) and
keeping terms of order $\varepsilon$ only, equation (\ref{nonl}) becomes
\be
2\,\rho\, v\, u_{0\xi \tau}+\fr{\pr}{\pr \xi}\int_R g(\xi-\xi^\prm)\,u_{0
\xi^\prm}\,d\xi^\prm+
\gamma\, u_{0 \xi}\,u_{0 \xi \xi}=-c_0^2\, u_{1xx}+\rho\,\, u_{1tt},
\label{olo}
\ee
where $c_0^2=\rho v^2$.

Then, the secularity condition  \cite{N} implies that both the left 
and right-hand  sides of (\ref{olo}) are zero. 
Thus, the right-hand side leads to the standard linear wave equation 
for $u_1$; while the left-hand side gives the long-time behaviour 
of $u_0$, \ie
\be
2\,\rho\, v\, u_{0\xi \tau}+\fr{\pr}{\pr \xi}\int_R g(\xi-\xi^\prm)\,u_{0
\xi^\prm}\,d\xi^\prm+
\gamma\, u_{0 \xi}\,u_{0 \xi \xi}=0.
\label{nonl1}
\ee
Finally, by letting 
\be
u_{0 \xi}={\cal F}, \hs \hs g(z)=\fr{\pr}{\pr z} G(z),
\label{g}
\ee
where $z=\xi-\xi^\prm$ and $y=\xi-z$, equation (\ref{nonl1}) becomes
\be
2\,\rho\,v\,{\cal F}_\tau+\gamma\,{\cal F}\,{\cal F}_\xi+\int_R 
G(\xi-y)\fr{\pr^2}{\pr^2 y}{\cal F}(y,\tau)\,dy=0,
\label{re}
\ee
which transforms to the Benjamin-Ono equation for
\be
\tau=\fr{1}{2\rho v}  T, \hs \hs {\cal F}=\fr{1}{\gamma}\, U, \hs \hs
G(\xi-y)=-\fr{\al}{\pi}\fr{1}{\xi-y}.
\label{G}
\ee

Note that, under the swap of the role played by $x$ and $t$ in (\ref{re}),
we arrive at an  integro-differential equation usually met in nonlinear
wave propagation in  active media with dissipative process \cite{Fr}. 
The latter is due to viscoelasticity described by an integral law 
involving exponential kernel in the integral part 
of the equation of motion. 
This kernel is then a function of time introduced by a relaxation process.

\subsection{The Benjamin-Ono Equation}

The Benjamin-Ono evolution equation
\be
U_T+U\,U_\xi+\alpha\,H(U_{\xi \xi})=0,
\label{U}
\ee
with $\al$ being a positive parameter, was originally derived in
\cite{B,DA,O}
for  interval-wave propagation in a two-layer system: one
shallow and the other infinitely deep.
In this context, $U(\xi, T)$ represents the amplitude of the interfacial wave
produced by an initial disturbance, say $U(\xi, 0)$.
The operator $H$ designates the one-dimensional Hilbert transform, \ie
\be
\alpha\,H(U_{\xi \xi})=\fr{\alpha}{\pi}\int_R \fr{1}{y-\xi}\, U_{yy}\, dy,
\label{BO}
\ee
which implies that (\ref{U}) is an integro-differential weakly nonlinear
evolution equation.

Equation (\ref{U}) has a simple solitary wave solution in the
form of a Lorentzian (algebraic) shape \cite{N,ABFS}, \ie
\be
U(\xi,T)=\fr{4\nu}{(\fr{\nu}{\al})^2(\xi-\nu T)^2+1},
\label{BOs}
\ee
where  $4\nu$ is the amplitude of the wave and $\fr{\al}{\nu}$ measures 
the wavelength.
The velocity of the soliton is amplitude-dependent and equals to $\nu$.

{\bf Remark:} 
Regarding the long-wavelength limit of the lattice model, the classical
 continuum approximation (\ie Taylor's expansion for the fields) of
(\ref{nonl}) leads to the local nonlinear partial
 derivative equation of the Boussinesq type \cite{PPF,FPR}
\be
u_{tt}-c^2u_{xx}+au_xu_{xx}+bu_{xxxx}=0,
\label{Bou}
\ee
where the parameters $c$, $a$ and $b$ depend on the atomic interactions. 
Alternatively, equation (\ref{Bou}) may be directly deduced as a special case
from the continuum approximation of (\ref{dis}).

The long-time evolution of  (\ref{Bou}) is given by the Korteweg-de-Vries 
equation, 
\be
V_t+V V_x+V_{xxx}=0,
\ee
when multi-scale techniques (or perturbation methods) as described 
in Section 3.2 are used.
In addition, the Benjamin-Ono equation (\ref{U}) can be transformed into the 
Korteweg-de-Vries equation by replacing $\alpha\,H(U_{\xi \xi})$ by 
$U_{\xi \xi \xi}$; however, this transition is not an obvious one.

\section{Evaluation of the Interaction Coefficients}

In this section we derive the lattice interaction 
coefficients $\Psi(n)$ of the model (\ref{ldis}) using 
the fact that its asymptotic expansion, for large times, leads to 
the Benjamin-Ono equation; and therefore, the kernel of the operator 
of elastic moduli $c(x)$ is known.

Since there is a correspondence between functions of discrete argument
 and a certain class of analytic functions, the interaction coefficients 
can be evaluated either in the space of functions of discrete argument
 or in the space of analytic functions.
However, since all these spaces are isomorphic to each other
the interaction coefficients are obtained using  
the $x$ representation (due to simplicity).

Equation (\ref{22}) by setting $x-x^\prime \ra x$ becomes
\begin{eqnarray}
\Phi(x)&=&-\fr{\pr^2}{\pr x^2}c(x)\nonumber\\
&=&-c_0^2 \delta^{\prm \prm}(x)-
\varepsilon\,g^{\prm \prm}(x),
\end{eqnarray}
where (\ref{c}) and (\ref{cod}) were used.
But from the long-time evolution analysis of the localized wave the
 explicit form of the function $g$ is known.
Thus, using (\ref{g}) and (\ref{G}) we get
\begin{eqnarray}
\Phi(x)&=&-c_0^2 \delta^{\prm \prm}(x)-\varepsilon\,G^{\prm \prm\prm}(x)
\nonumber\\
&=&-c_0^2 \delta^{\prm \prm}(x)-\fr{6\varepsilon \al}{\pi}\fr{1}{x^4}.
\end{eqnarray}

Due to Fourier transform, the force constants in the space of 
discrete argument are related with the ones in the space of 
analytic functions as \cite{K}:
\begin{eqnarray}
\Phi(n)=\Phi(nb)&=&\int_R \Phi(x)\, \delta_b(x-nb)\, dx\nonumber\\
&=&-c_0^2\,{\cal I}_1(n)-\fr{6\al \varepsilon}{\pi}\,{\cal I}_2(n),
\end{eqnarray}
where $\delta_B (x) \doteq (2\pi)^{-1}\int_B e^{ikx} dx
=\sin(\pi x)(\pi x)^{-1}$
 and
\begin{eqnarray}
{\cal I}_1(n)&=&\int_R \delta^{\prm \prm}(x) \fr{\sin[\fr{\pi}{b}(x-nb)]}
{\pi (x-nb)}\,dx,\nonumber\acc
{\cal I}_2(n)&=&\int_R\fr{1}{x^4}
 \fr{\sin[\fr{\pi}{b}(x-nb)]}{\pi (x-nb)}\,dx.
\end{eqnarray}

Thus, it is a matter of algebra to evaluate the two integrals and obtain
\begin{eqnarray}
{\cal I}_1(0)&=&-\fr{\pi^2}{3\,b^3},\nonumber\\
{\cal I}_2(0)&=&\fr{\pi^4}{24\,b^4},
\end{eqnarray}
while for $n \neq 0$
\begin{eqnarray}
{\cal I}_1(n)&=&-\fr{2}{b^3}\fr{(-1)^n}{n^2},\nonumber\acc
{\cal I}_2(n)&=&\fr{\pi^2}{b^4}\fr{(-1)^n}{n^2}\left[\fr{1}{2}
+\fr{(-1)^n-1}{n^2\,\pi^2}\right].
\end{eqnarray}

Therefore, the lattice force constants are given by
\begin{eqnarray}
\Phi(0)&=&\fr{c_0^2 \pi^2}{3b^3}-\fr{\al\varepsilon\pi^3}{4b^4},\nonumber\\
\Phi(n)&=&\fr{2c_0^2}{b^3}\fr{(-1)^n}{n^2}-\fr{6\al\varepsilon 
\pi}{b^4}\fr{(-1)^n}
{n^2}\left[\fr{1}{2}+\fr{(-1)^n-1}{n^2\,\pi^2}\right],\hs n \neq 0.
\label{morfi}
\end{eqnarray}

Finally, the lattice interaction coefficients $\Psi(n)$ in (\ref{dis}) 
(due to (\ref{Pdis})) are given by
\be
\Psi(n)=\Phi(0)\delta(n)-\Phi(n).
\label{psi}
\ee
As one can observe from Figure 2, by keeping the nonlocality of the discrete
model to the continuum one, and specifically, the choice of an 
integrable nonlocal equation as its asymptotic limit for large times, 
leads to the construction of a lattice model with negative interaction 
coefficients between its particles.
In fact, Figure 2 illustrates the change of sign of $\Psi(n)$
and its variation with $n$ for an atomic chain where each particle interacts
with its ten neighbour particles.
[The values of the other parameters, \ie $(a,b,\varepsilon, c_0)$ are
 listed in the figure caption.]

\begin{figure}
\unitlength1cm
\begin{center}
\epsfxsize=8cm
\epsffile{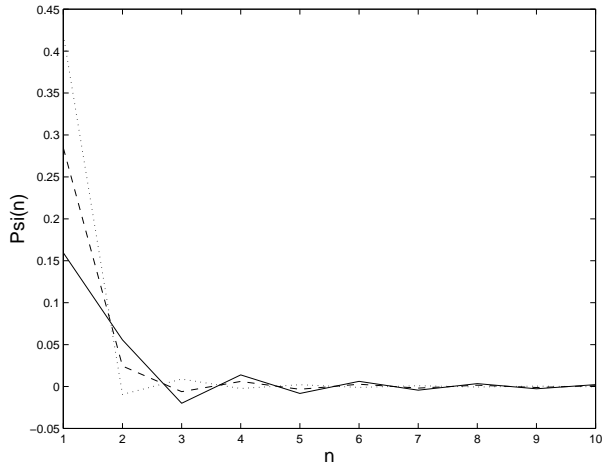}
\end{center}
\caption{The function $\Psi(n)$ given by (\ref{morfi})-(\ref{psi}) for 
$\alpha=1$, $\nu=1/4$, $\varepsilon=0.1$, $n_0=125$ and phase 
velocity (a) $c_0=0.6$ solid line (b) $c_0=0.65$ dashed line 
(c) $c_0=0.7$  dotted line.}
\end{figure}

\section{Numerical Simulations}

The behaviour of genuinely discrete systems is quite different than the
one of their continuum counterparts and these differences have been studied
in numerous papers both from the mathematical and the physical point of
view.
Many of the methods used and their physical implementation  in models with
nearest neighbour interactions can be found in \cite{Bal,Bal1,Bal2}.
In addition, in \cite{Bal3,JA} the effects of discreteness in solitary 
waves in nearest neighbours lattices have been mathematically 
investigated.

In our case, the lattice model keeps its nonlocal behaviour at the continuum
when its asymptotical limit for large times is the Benjamin-Ono equation. 
However, it is not obvious if the solution of the Benjamin-Ono equation 
will be the solution of the lattice model, as well. 
Thus, in this section we investigate the dynamical behaviour of the
lattice model (\ref{dis}) numerically, using as an initial condition
the soliton solution given in Section 3.3. 

We propose a numerical scheme by directly
considering the lattice equations (\ref{dis}), employing
a Runge-Kutta method of fourth order and imposing
pseudo-periodic boundary conditions, \ie
\be
u(N+i)=u(i)+u_0, \hs \hs \hs i \in [-{\cal M},{\cal M}],
\ee
where $N$ is the number of particles in the lattice,  
${\cal M}$ is the range of particle interactions considered and $u_0$ is the
amplitude of the initial kink.

The initial conditions for the displacements and the velocities of each
lattice particle are given by the analytic expression 
\be
u(x,t)=\fr{2 \varepsilon \al}{k_3}
\arctan\!\!\left[\fr{\nu}{\al}\left(x-x_0-c_0\left(1+\fr{\varepsilon
\nu}{2c_0^2}\right)t\right)\right],
\label{IC}
\ee
and its time derivative, respectively.
Here $x_0$ denotes the soliton position and we have set the mass density 
equal to one ($\rho=1$).
This solution represents a kink  with thickness 
$\Delta=\alpha \pi/\nu$, \ie the number of particles participating in the
kink solution is proportional to the kernel factor $\alpha$ 
and decreases as the wave parameter $\nu$ increases.
[The solution (\ref{IC}) has been obtained from the solitary wave 
solution of the Benjamin-Ono equation (\ref{BOs})  after using the 
change of variables  and  fields given in Section 3.2].
Figure 3 illustrates (a) the soliton solution $U(\zeta)$ with
$\zeta=\xi-\nu T$ given by  (\ref{BOs}) and (b) the kink solution $u(x)$
 given by (\ref{IC}) for the choice $b=1$.

\begin{figure}
\unitlength1cm 
\begin{picture}(5,5)
\put(.5,5.5){(a)}
\epsfxsize=6.5cm   
\epsffile{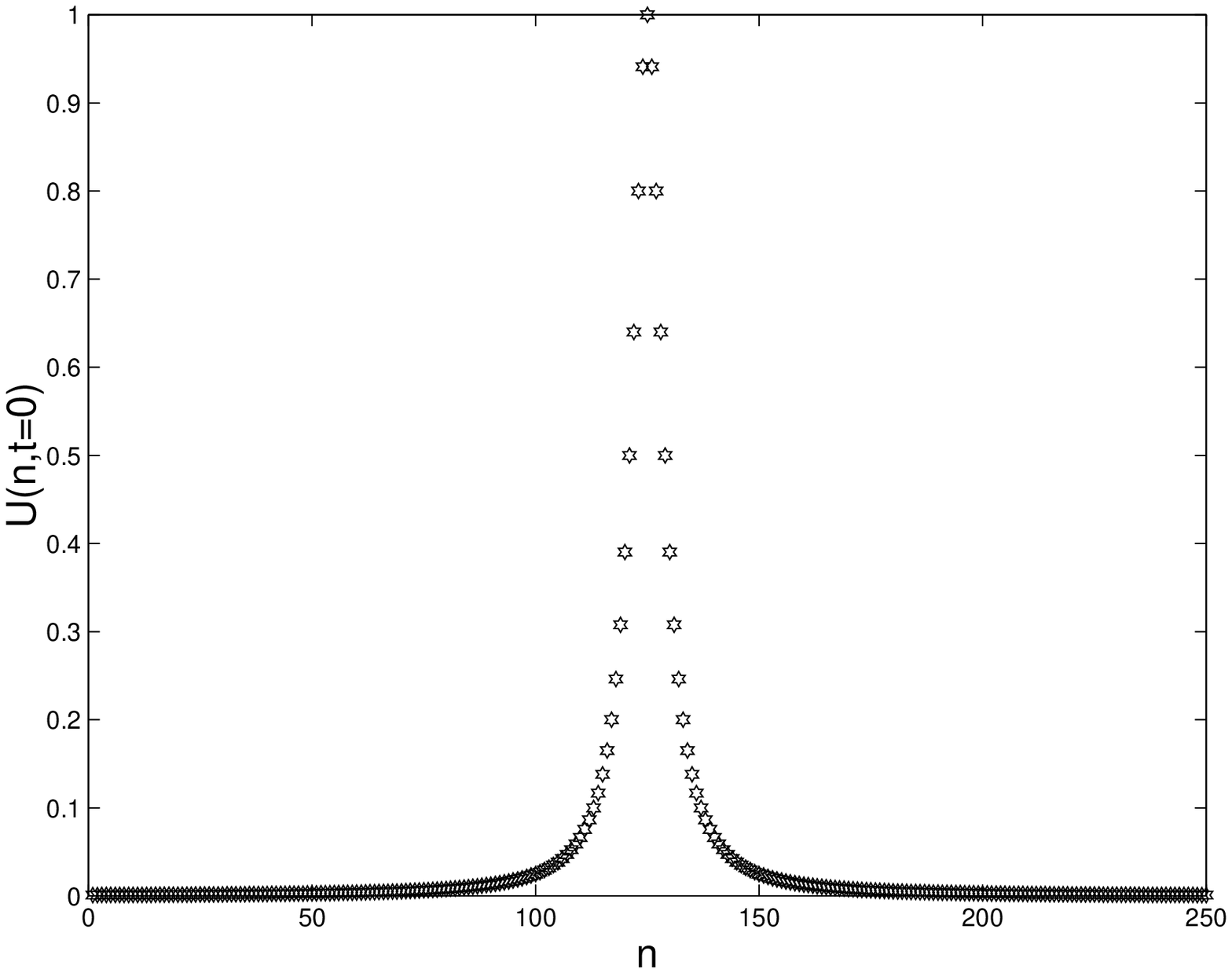}
\end{picture}
\hfill
\begin{picture}(8,0)
\put(.5,5.5){(b)}
\epsfxsize=6.5cm
\epsffile{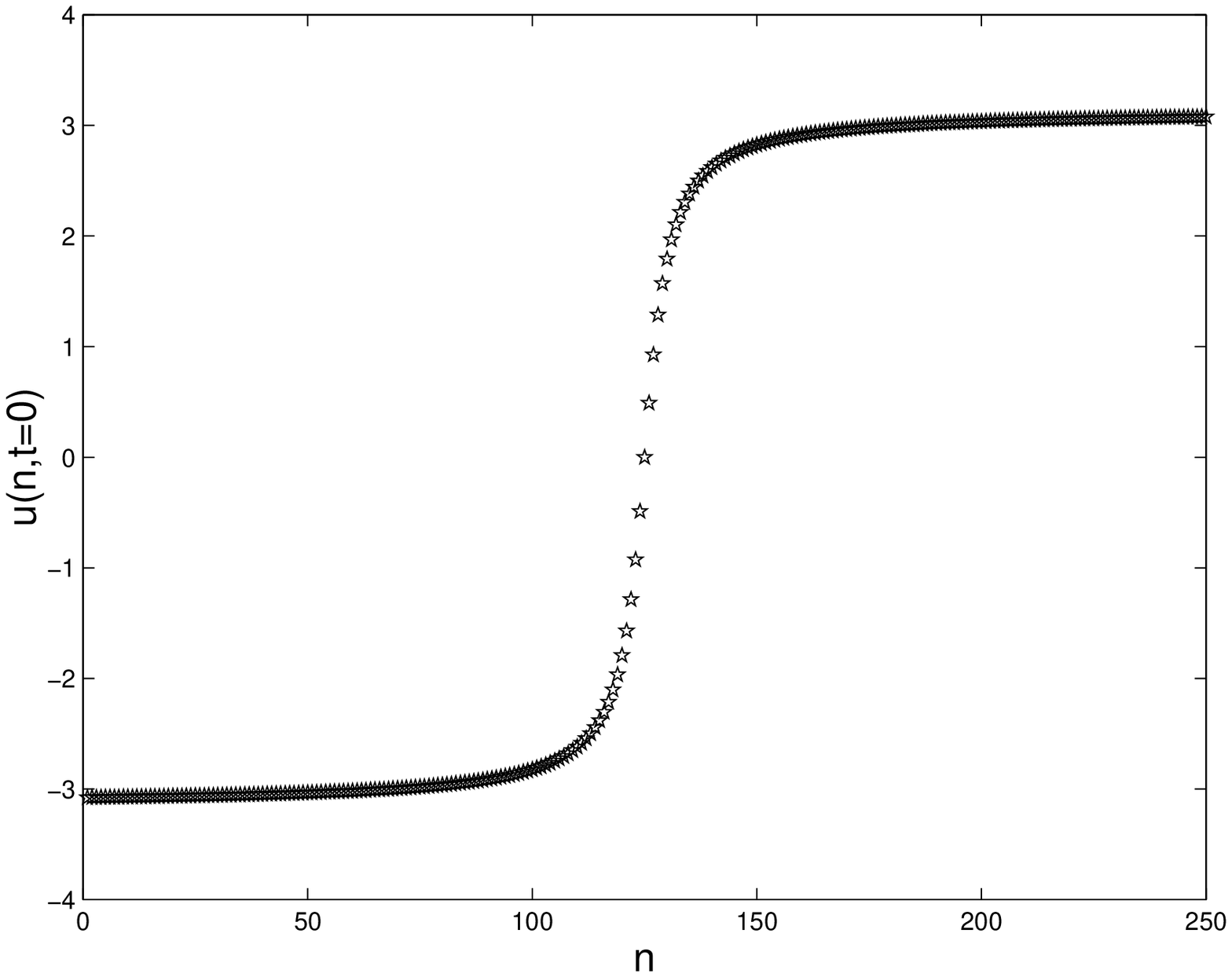}
\end{picture}
\caption{(a) Initial soliton solution (\ref{BOs}) and (b) 
Initial kink solution (\ref{IC}) for $\alpha=1$, $\nu=1/4$ and
 $n_0=125$.}
\end{figure}

In all our simulations  we have chosen the following fixed values 
for: $k_3=0.1$, $\varepsilon=0.1$, $b=1$, $\al=1$, $M=1$.
However, different values for the above parameters give qualitatively the
same results.
The initial position of the kink is chosen equal to $n_0=125$,
 while the value of the kink velocity $c_0$ is chosen so that
the first interaction coefficient is positive (\ie  $\Psi(1)>0$) and
the total number of particles  in the lattice is equal to 250.

Figure 4 summarises the main results of our numerical simulations when
 models with different number of non-nearest neighbour 
interactions are considered and different values for the velocity and
 wavelength  of the initial kink have been chosen.
In particular, Figures 4(a), 4(e) and 4(f) present a solution 
corresponding to  a lattice model with  first, tenth and 
twentieth neighbour particle interactions  when the initial kink
wavelength and phase velocity are: $\nu=1/8$, $c_0=0.7$;  $\nu=1/4$,
$c_0=0.6$; and $\nu=1/8$, $c_0=0.6$, respectively.
In each model, the kink contains from twelve up to twenty four particles, 
which means that the continuum approximation is valid. 
[Recall that the kink thickness depends on the parameter $\nu$, due to
 (\ref{IC})].
On the other hand, Figures 4(b), 4(c) and 4(d) present the kink dynamics
 for lattice models with second, fourth and seventh neighbour particle
 interactions. 
However, we choose the kink width in all three cases to be small,
\ie the kink contains six particles with $\nu=1/2$, $c_0=0.54$ in the
first case, and three particles with $\nu=1$; $c_0=0.6$, $c_0=0.7$ 
in the last two.

From Figure 4, one can observe that the initial soliton solution of the 
Benjamin-Ono equation is also a solution of the lattice model
with long-range interactions given by (\ref{dis}).
In fact, the initial lattice kink relaxes and propagates in the lattice space
with small oscillations.
It can be observed from Figures  4(a), 4(e) and 4(f) 
that as the kink thickness increases (\ie $\nu$ decreases) 
the lattice radiation  is eliminated.
Numerical simulations of the discrete nonlocal model for kink widths 
equal to three or six lattice spacings have been presented 
in Figures 4(b)-4(d): 
at this level the kink changes slightly and loses
energy by radiation of small amplitudes waves.
In all  cases where the kink contains at least 20 particles, 
it is remarkable stable and emits (almost) no radiation.

Let us conclude by saying that, in the long-wavelength approximation (ie, when
the average kink wavelength is larger compared to the lattice spacing) the 
initial (continuum) kink, which has been derived  using the quasi-continuum 
approximation of the discrete model, relaxes  to a stable discrete kink
without any oscillations and our method is accurate.

\begin{figure}
\unitlength1cm
\begin{picture}(5,5)
\put(1,5.5){(a)}
\epsfxsize=6.5cm
\epsffile{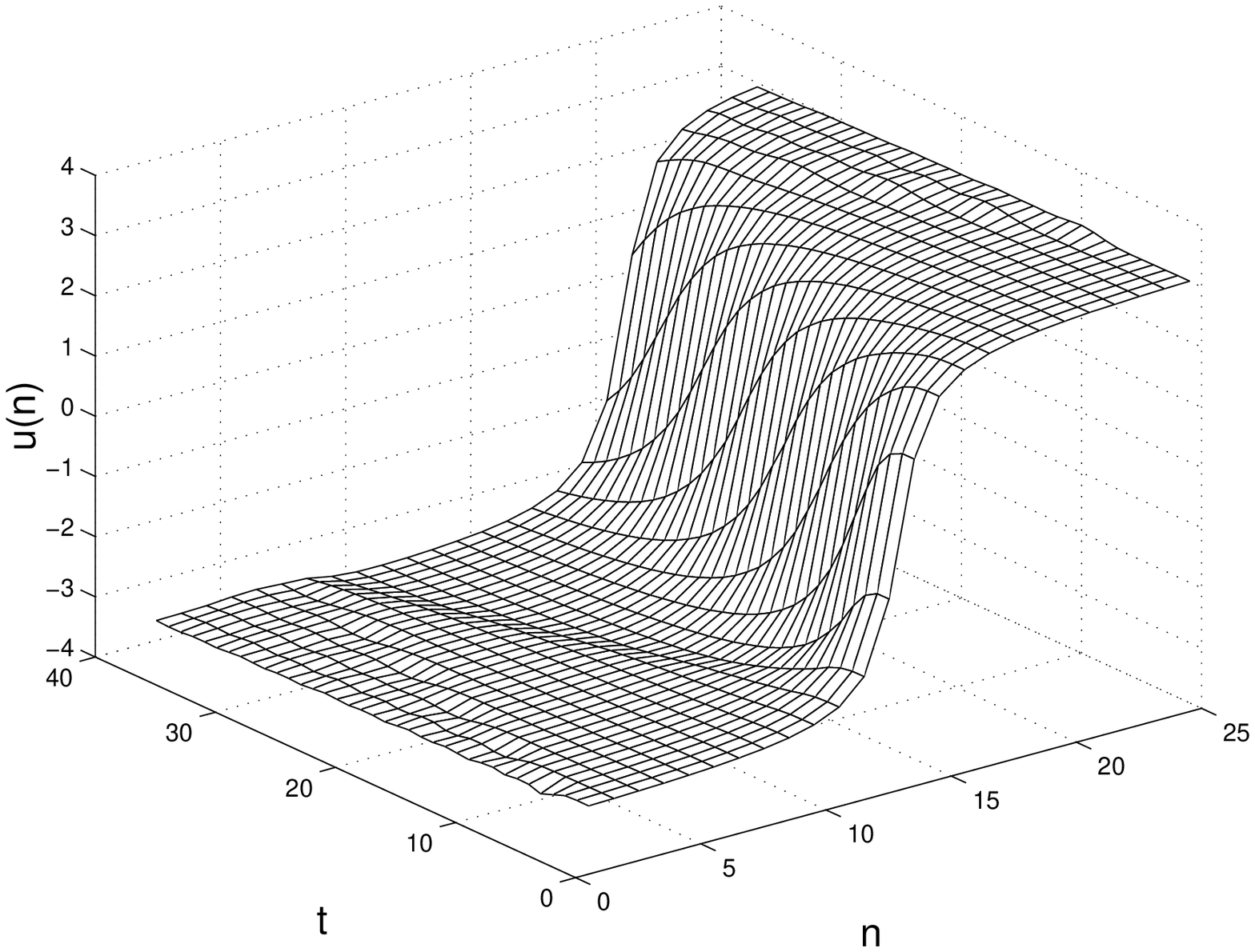}
\end{picture}
\par
\hfill
\begin{picture}(8,0)
\put(1,6){(b)}
\epsfxsize=6.5cm
\epsffile{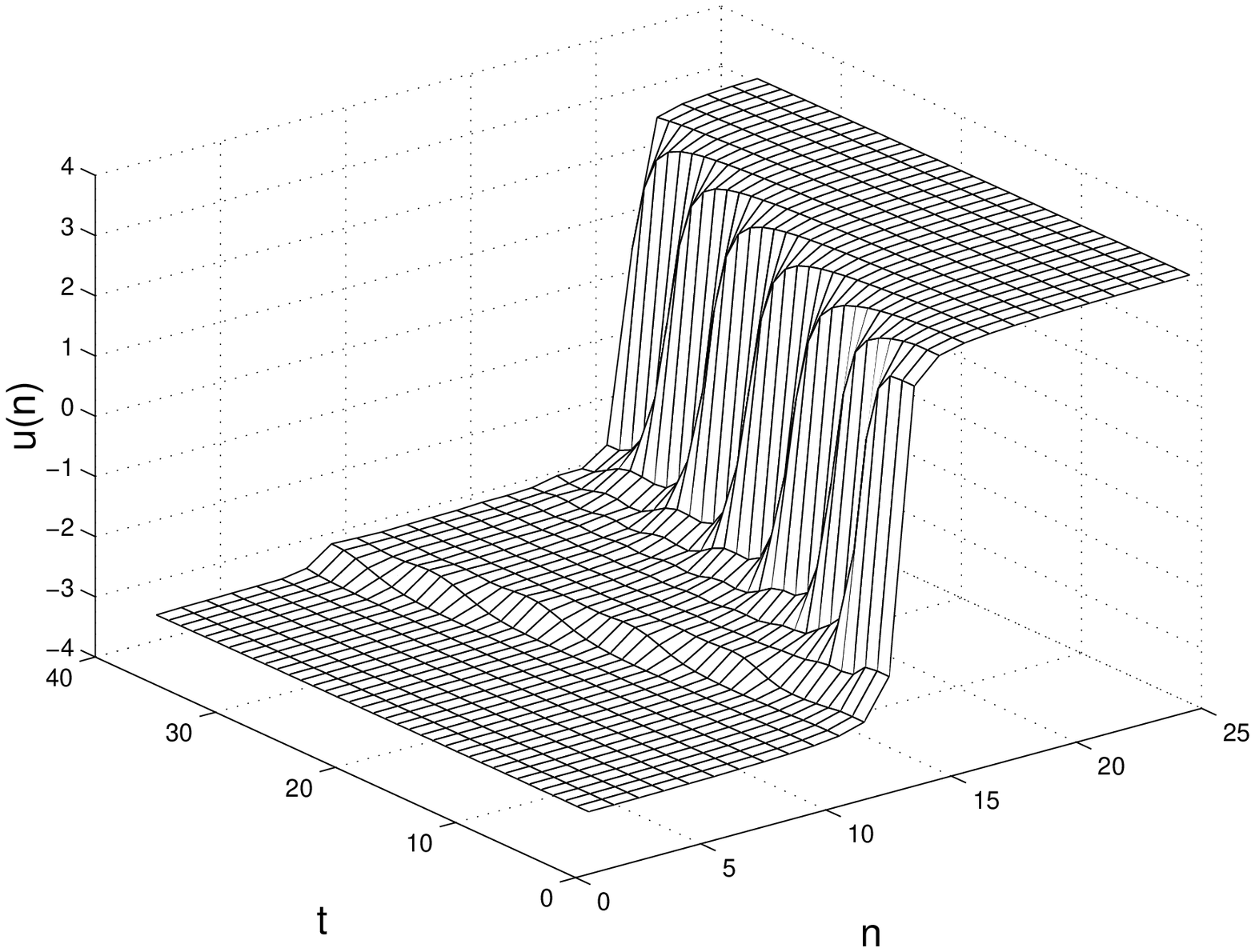}
\end{picture}
\par
\hfill
\begin{picture}(17,6)
\put(1,5.5){(c)}
\epsfxsize=6.5cm
\epsffile{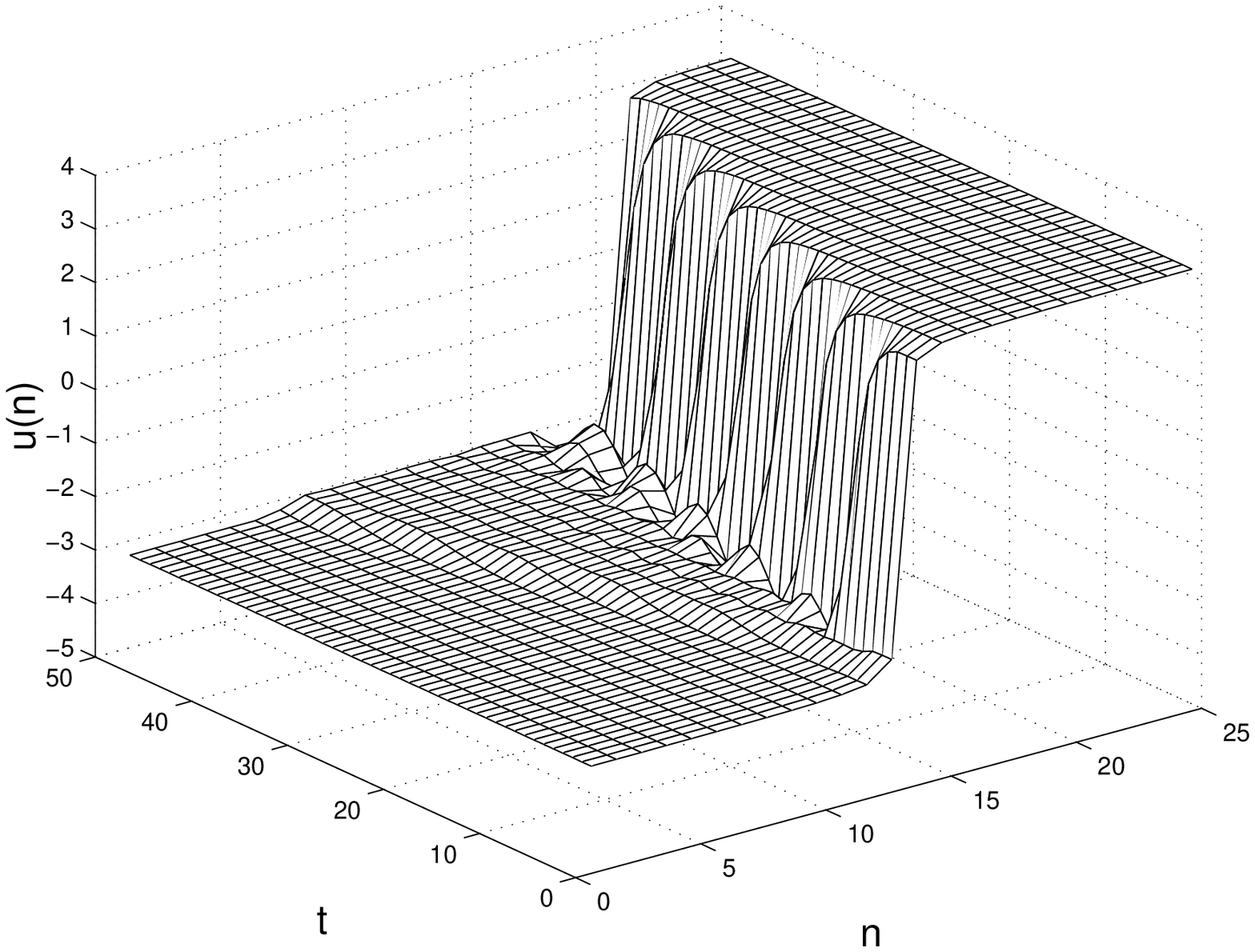}
\end{picture}
\par
\hfill
\begin{picture}(8,0)
\put(1,6){(d)}
\epsfxsize=6.5cm
\epsffile{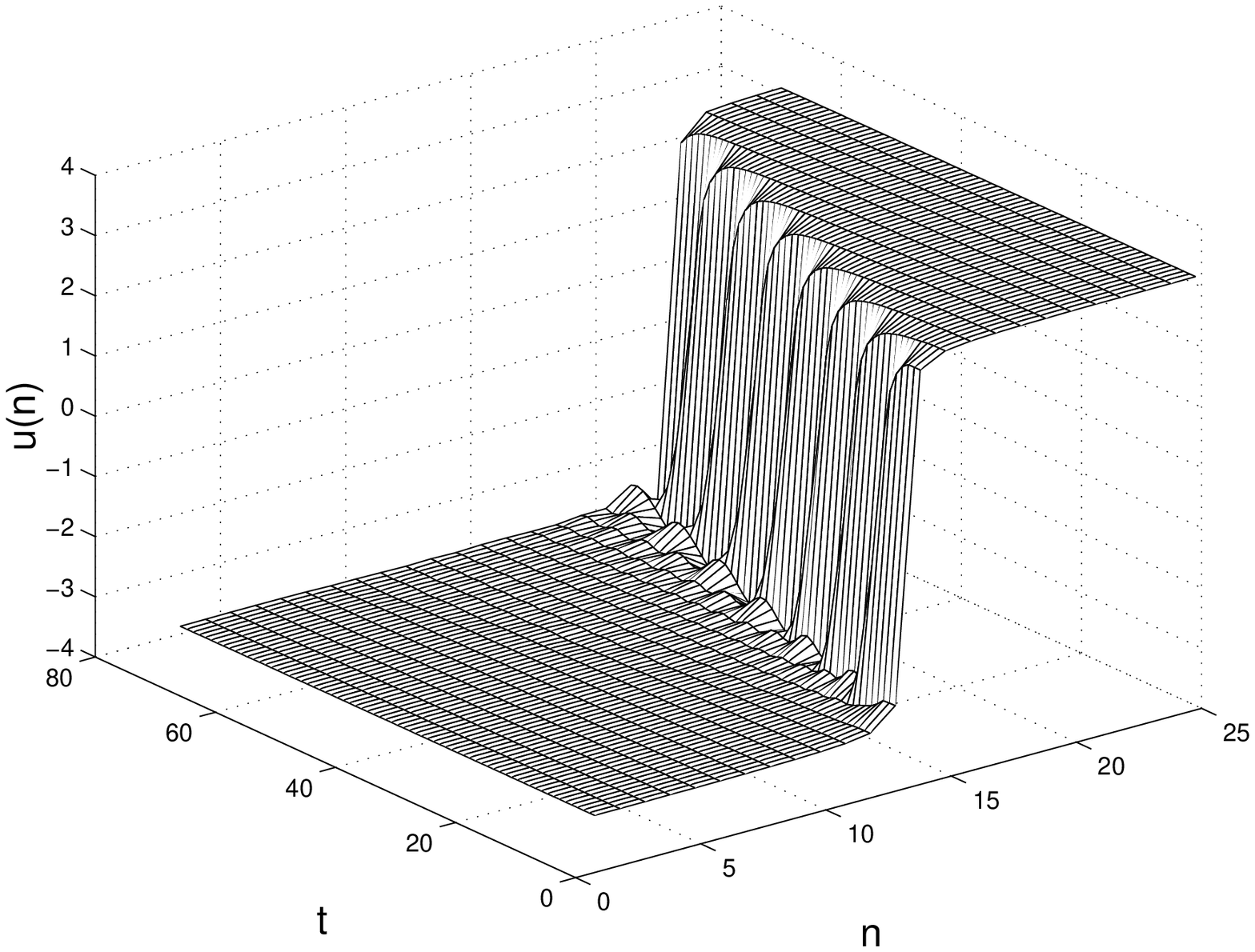}
\end{picture}
\begin{picture}(17,6)
\put(1,5.5){(e)}
\epsfxsize=6.5cm
\epsffile{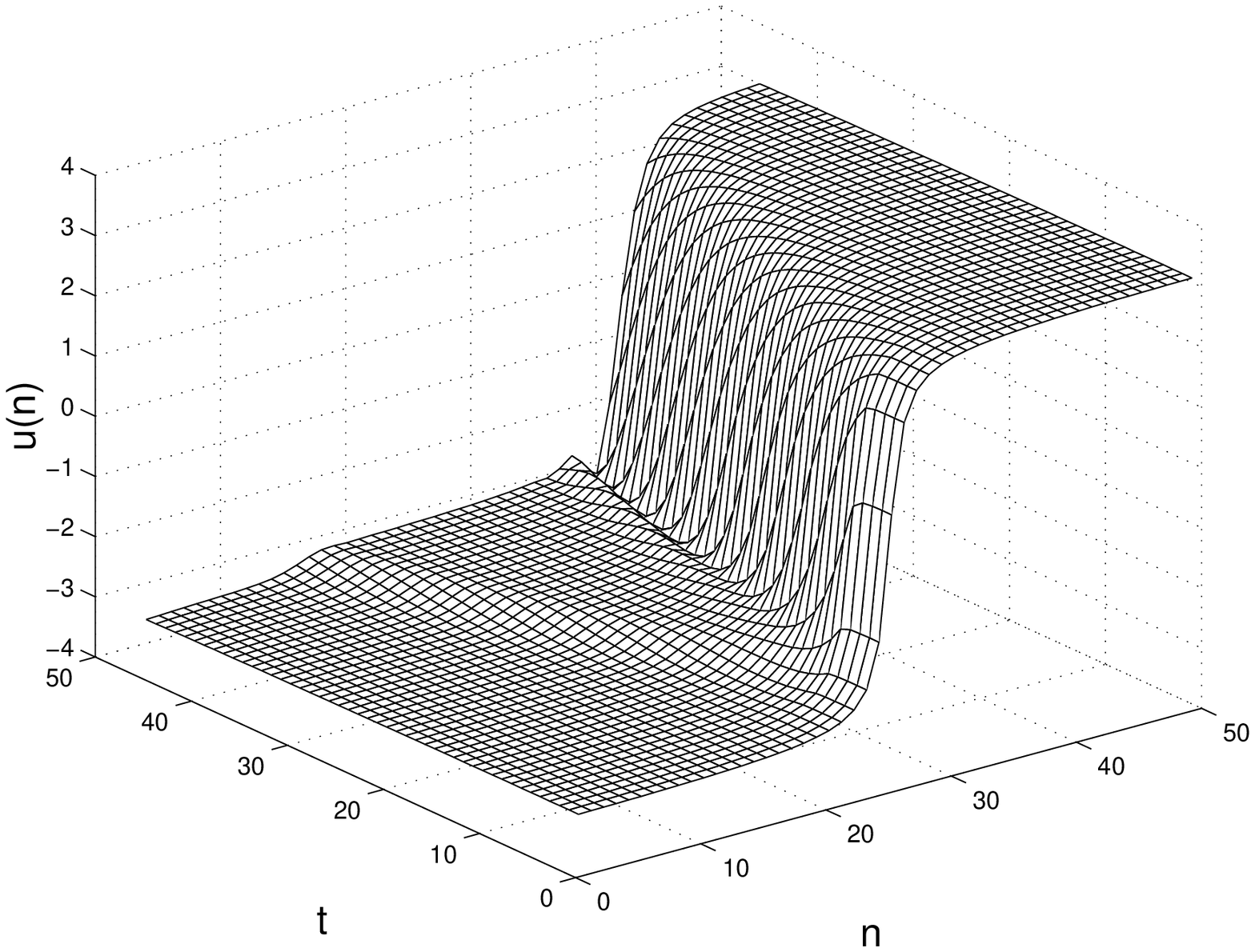}
\end{picture}
\par
\hfill
\begin{picture}(8,0)
\put(1,6){(f)}
\epsfxsize=6.5cm
\epsffile{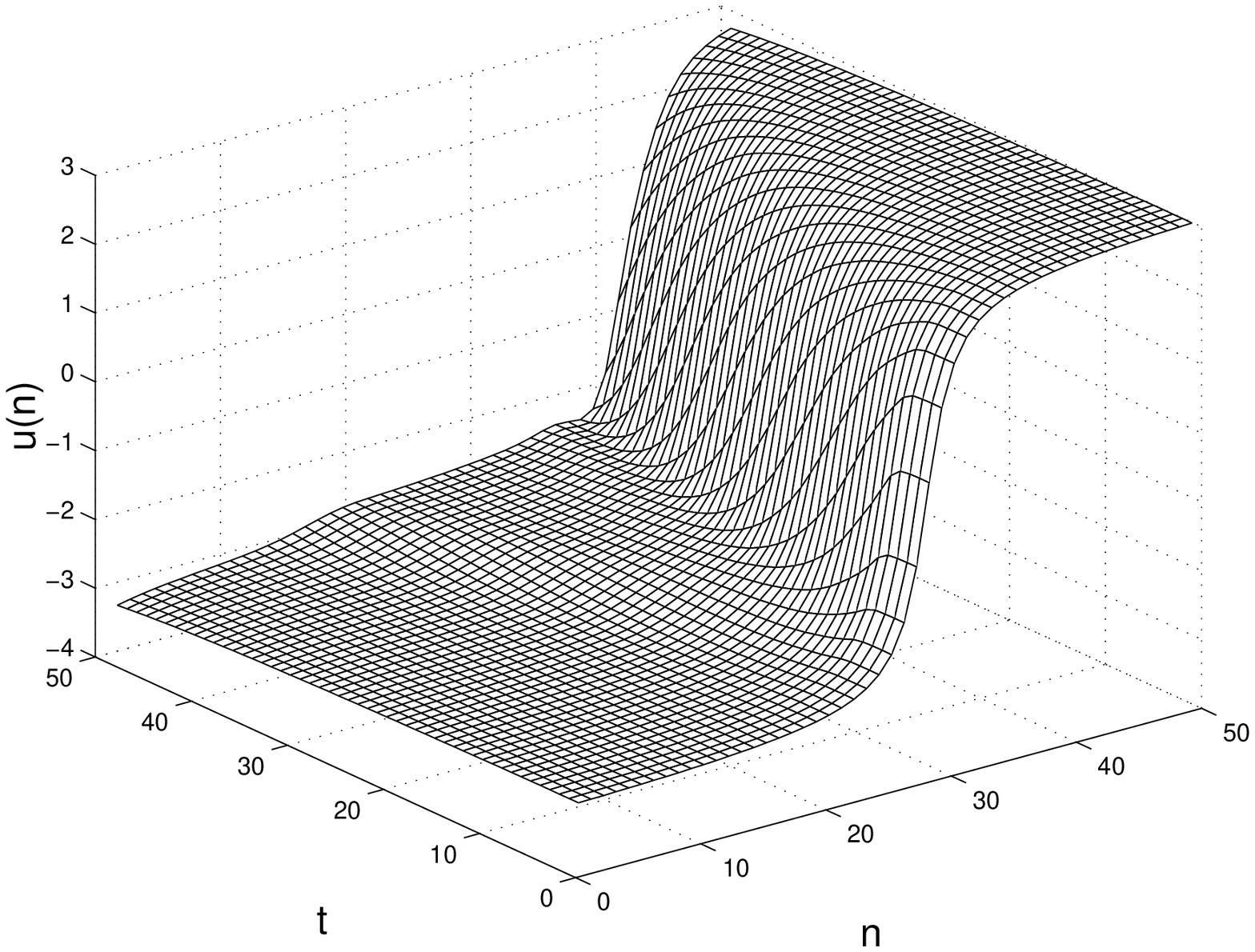}
\end{picture}
\caption{Kink plane for (a) one-range interaction,  $\nu=1/8$, $c_0=0.7$
(b) two-range interaction,  $\nu=1/2$, $c_0=0.54$
(c) four-range interaction,  $\nu=1$, $c_0=0.6$
(d) seven-range interaction,  $\nu=1$, $c_0=0.7$
(e) ten-range interaction, $\nu=1/4$, $c_0=0.6$
(f) twenty-range interaction,  $\nu=1/8$, $c_0=0.6$.}
\end{figure}

\section{Concluding Remarks}

The present work is concerned with the dynamics of
lattice models with long-range and nonlinear interactions.
The main objective is the description of the dynamics of the lattice model,
 at the
continuum level,  by keeping the nonlocal nature of the discrete system.
Our model is the simplest one consisting of a one-dimensional chain of point
masses connected by springs.
We have developed a theory for such a discrete system and introduced
 the notion of
quasicontinuum which allows us to treat the discrete and continuum models in 
the framework of the same formalism.
Attention has been focused to the analysis of an approximate model 
and its transition to nonlocal elasticity in the limit of long waves.
In this formalism, we have set up a one-to-one correspondence between
functions of discrete arguments and a class of analytic functions.
The procedure is also applicable for integral operations on these functions.
The method brings out a resemblance to the problem of interpolation of a
discrete function by a smooth function satisfying some smoothness conditions.
The technique allows us (Section 3) to use the same representation of the
Lagrangian  for all cases: discrete and analytic.

The equation of motion of the model in the continuum representation is
similar to
that of a nonlocal model of elasticity including a local nonlinear term.
Nevertheless, such an equation of motion is not easy to deal with and a
perturbation method has been employed.
More precisely, we have considered the long-time evolution of
 nonlinear
waves traveling in the lattice.
The asymptotic method has been used to transform the equation of motion into 
an equation of the Benjamin-Ono type  which governs the long-time
evolution of the initial signal.
For a particular choice of the nonlocal elastic kernel or interaction force
function, the dispersive operator of the equation is a Hilbert transform
corresponding to the Benjamin-Ono equation.
In this particular situation, a localized wave solution of soliton type has
 been obtained.
In order to test the analytical conjectures thus developed,
 some numerical simulations have been performed directly on
 the lattice model for different ranges of  nonlocal action.
It has been observed from numerics that the initial soliton (kink) solution 
propagates almost without perturbation and it is very stable.
Moreover, wide kinks propagate in the lattice without any loss due to
discreteness radiations.

Such a study can be extended to other kinds of nonlinear terms (double-well 
or sine-potential, higher-order interactions) or to arbitrary particle
interactions. 
Extension to two-dimensional lattices can  also be envisaged.
The problems of kink collision or the influence of perturbations (forces,
damping, lattice defects) on the kink motion is of interest.
Some of these problems will be addressed in the future.

\section*{Acknowledgements}
This work has been performed in the framework of the TMR European Contract 
{\bf FMRX-CT-960062:} {\it Spatio-temporal instabilities in
deformation and fracture mechanics, material science and nonlinear physics
aspects}.
TI acknowledges the Nuffield Foundation for a newly appointed lecturer award.

\end{document}